\documentclass[aps,twocolumn,prd,showpacs,showkeys,preprintnumbers,superscriptaddress,bibnotes,floatfix,longbibliography]{revtex4-1}

\pdfoutput=1
\usepackage{amsmath}
\usepackage{amsfonts}
\usepackage{amssymb}
\usepackage{mathrsfs}
\usepackage{color}
\usepackage{bm}
\usepackage{graphicx}
\usepackage{blindtext}
\usepackage{wasysym}
\usepackage{mwe}
\usepackage{hyperref}
\usepackage[normalem]{ulem}

\newcommand{\ie}{{\it i.e.}}

\newcommand{\eg}{{\it e.g.}}

\newcommand{\fig}{Fig.}

\newcommand{\Ref}{Ref.}
\newcommand{\Refs}{Refs.}

\newcommand{\figu}[1]{\fig~\ref{fig:#1}}

\newcommand{\bi}{\begin{itemize}}
\newcommand{\ei}{\end{itemize}}

\newcommand{\matr}[1]{\mathbf{#1}} 

\begin{document}
 
\title{Inferring the flavor of high-energy astrophysical neutrinos at their sources}
\author{Mauricio Bustamante}
\email{mbustamante@nbi.ku.dk}
\thanks{ORCID: \href{http://orcid.org/0000-0001-6923-0865}{0000-0001-6923-0865}}
\affiliation{Niels Bohr International Academy \& Discovery Centre, Niels Bohr Institute,\\University of Copenhagen, DK-2100 Copenhagen, Denmark}
\affiliation{DARK, Niels Bohr Institute, University of Copenhagen, DK-2100 Copenhagen, Denmark}

\author{Markus Ahlers}
\email{markus.ahlers@nbi.ku.dk}
\thanks{ORCID: \href{http://orcid.org/0000-0003-0709-5631}{0000-0003-0709-5631}}
\affiliation{Niels Bohr International Academy \& Discovery Centre, Niels Bohr Institute,\\University of Copenhagen, DK-2100 Copenhagen, Denmark}

\date{June 21, 2019}

\begin{abstract}

The sources and production mechanisms of high-energy astrophysical neutrinos are largely unknown.  
A promising opportunity for progress lies in the study of neutrino flavor composition, \ie, the proportion of each flavor in the flux of neutrinos, which reflects the physical conditions at the sources.  To seize it, we introduce a Bayesian method that infers the flavor composition at the neutrino sources based on the flavor composition measured at Earth.  We find that present data from the IceCube neutrino telescope favor neutrino production via the decay of high-energy pions and rule out production via the decay of neutrons.  In the future, improved measurements of flavor composition and mixing parameters may single out the production mechanism with high significance.

\end{abstract}


\maketitle


{\bf Introduction.---}  High-energy astrophysical neutrinos with TeV--PeV energies provide crucial and unique information to understand the non-thermal Universe\ \cite{Anchordoqui:2013dnh, Ahlers:2018fkn}.   Yet, though firmly detected\ \cite{Aartsen:2013bka,Aartsen:2013jdh,Aartsen:2014gkd,Aartsen:2015rwa,Aartsen:2016xlq}, they have a largely unknown origin.  They likely come predominantly from extragalactic sources\ \cite{Ahlers:2013xia, Ahlers:2015moa, Denton:2017csz, Aartsen:2017ujz, Ahlers:2018fkn}, but, to date, no point-like source is known with certainty, notwithstanding noteworthy recent findings\ \cite{IceCube:2018dnn,IceCube:2018cha}.
In the future, improved event statistics, reduced systematic uncertainties, and combined multi-messenger analyses
will boost the prospects of discovering sources\ \cite{Aartsen:2014njl,Adrian-Martinez:2016fdl}. 

A complementary opportunity for progress, accessible already today, lies in measuring the flavor composition of high-energy astrophysical neutrinos, \ie, the relative number of neutrinos of each flavor.  The flavor composition that neutrinos are emitted with is the result of production processes that depend on the physical conditions in the astrophysical sources.  Therefore, flavor measurements can help to discriminate between candidate source classes\ \cite{Barenboim:2003jm, Xing:2006uk, Pakvasa:2007dc, Lai:2009ke, Choubey:2009jq}.
After emission, as neutrinos propagate {\it en route} to Earth, flavor oscillations modify the composition.  Neutrino telescopes, like IceCube, measure the flavor composition of the arriving flux.  Based on it, one can, in principle, infer the composition at the sources. 

Yet, existing analyses are either largely focused on inferring the flavor composition at Earth from data\ \cite{Mena:2014sja, Aartsen:2015ivb,Aartsen:2015knd, Palladino:2015vna, Palomares-Ruiz:2015mka, Vincent:2016nut,Aartsen:2018vez} or confined to assessing the compatibility of the flavor composition measured at Earth with expectations from a few idealized scenarios of neutrino production.   We follow an alternative strategy, hitherto unexplored, that provides more insight.  Using Bayesian statistics, we infer the composition at the sources based on the composition measured at neutrino telescopes, accounting for the uncertainties in its measurement and in the neutrino mixing parameters that drive oscillations.

\begin{figure}[t!]
 \centering
 \includegraphics[width=\columnwidth]{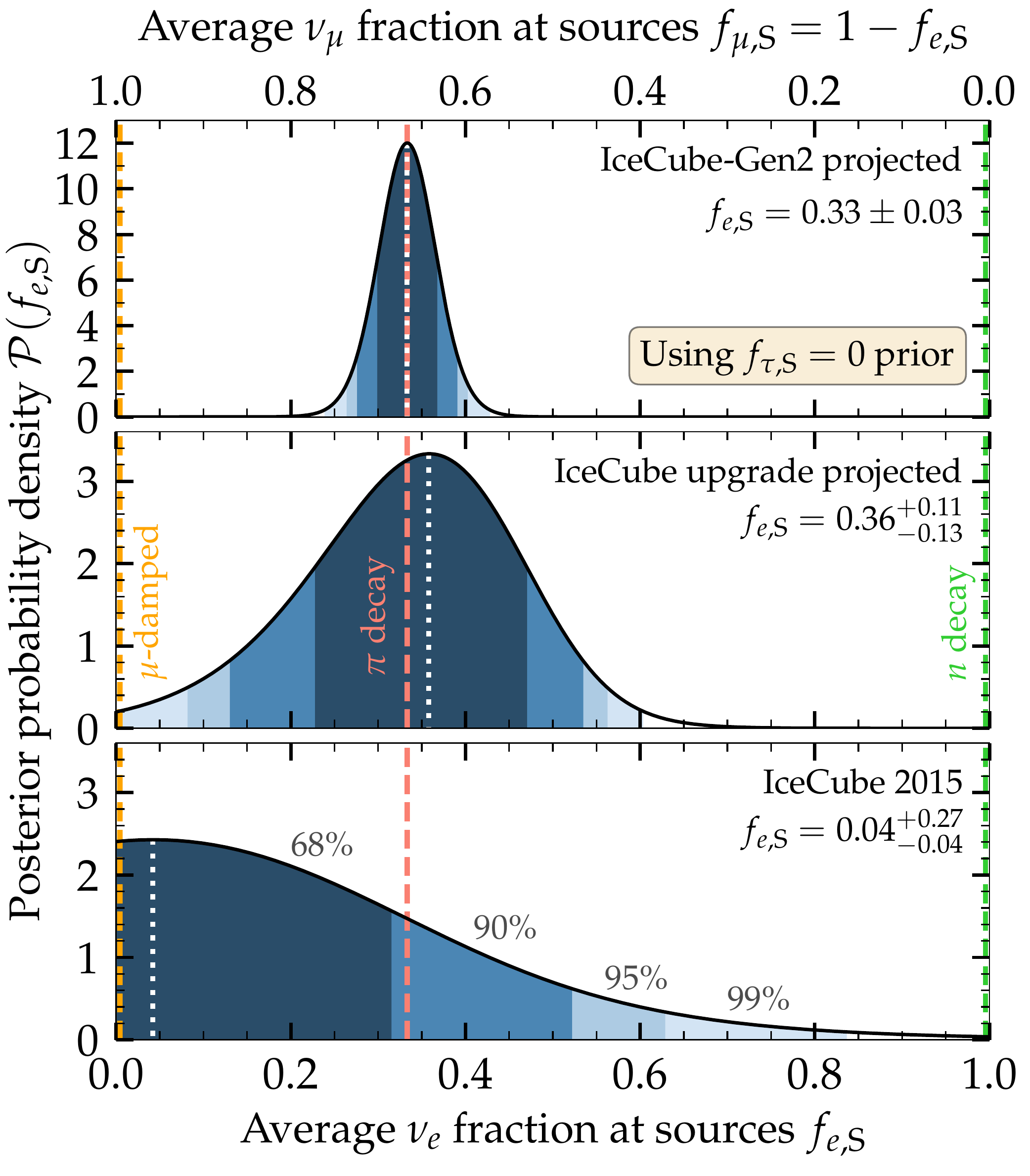}
 \caption{\label{fig:posterior_source_no_tau}Flavor composition of high-energy astrophysical neutrinos at their sources, inferred from present IceCube measurements\ \cite{Aartsen:2015knd} ({\it bottom}) and from the projected sensitivities of the near-future IceCube upgrade\ \cite{blot_talk_tevpa_2018} ({\it center}) and planned IceCube-Gen2\ \cite{kowalski_NOW_2018} ({\it top}), assuming production by pion decay.  Here we enforce a prior of no $\nu_\tau$ production, \ie, $f_{\tau, {\rm S}} = 0$.  We show the most probable values (white dotted lines) and credible intervals (blue shaded regions) of $f_{e, {\rm S}}$, and mark physically motivated neutrino production scenarios: pion decay, muon-damped, and neutron decay.}
\end{figure}

Figure \ref{fig:posterior_source_no_tau} shows our results obtained using published and projected flavor measurements in IceCube.  We report results in terms of flavor ratios $f_{\alpha,{\rm S}}$ ($\alpha = e, \mu, \tau$), \ie, the relative contribution of $\nu_\alpha + \bar{\nu}_\alpha$ to the total emission.  We find that present data favor neutrino production via the decay of high-energy pions and the synchrotron-cooling of intermediate muons in strong magnetic fields, and strongly disfavor production via neutron decay.


\begin{figure*}[t!]
 \centering
 \includegraphics[width=\columnwidth]{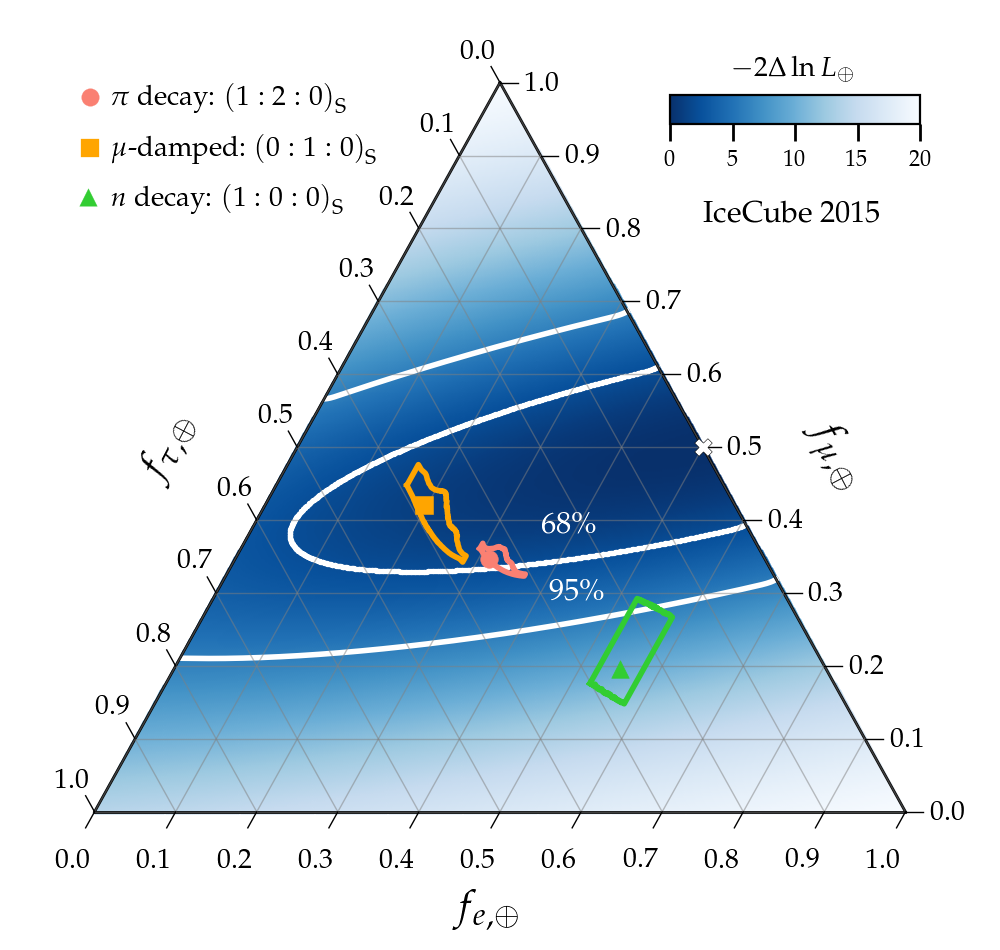}
 \includegraphics[width=\columnwidth]{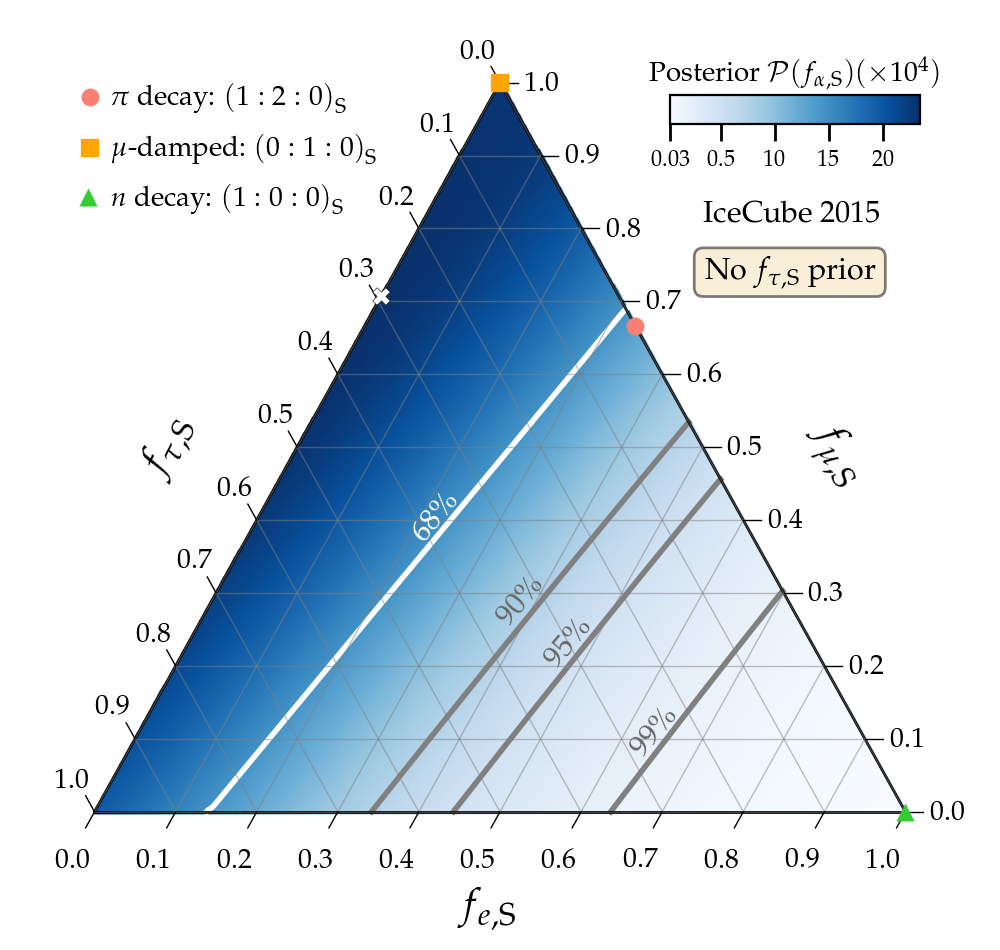}
 \caption{\label{fig:ic_2015} {\it Left:}  Flavor composition of high-energy astrophysical neutrinos at Earth, approximating current IceCube measurements\ \cite{Aartsen:2015knd}, expressed in terms of variations in the likelihood, $-2 \Delta \ln \mathcal{L}_\oplus$.  The contours show the 68\% and 95\% confidence regions; this triangle was produced by the IceCube Collaboration using a frequentist approach.  We include expectations from three benchmark production scenarios, computed with mixing parameters fixed at their best-fit values --- shown as symbols --- and varied within their $3\sigma$ ranges \cite{Esteban:2016qun, NuFit_Jan2018} --- shown as bounded regions. {\it Right:}  Flavor composition at the neutrino sources, inferred based on current measurements of flavor in IceCube and of mixing parameters in oscillation experiments\ \cite{Esteban:2016qun, NuFit_Jan2018}.  We assume no prior on the flavor composition at the sources.  The contours show the 68\%, 90\%, 95\%, and 99\% credible regions; this triangle was produced by the procedure introduced here using a Bayesian approach.}
\end{figure*}

\smallskip
{\bf Producing astrophysical neutrinos.---}  We expect astrophysical sources of high-energy neutrinos to accelerate protons beyond PeV energies via collisionless shocks in magnetized environments.  High-energy protons interact with ambient matter\ \cite{Margolis:1977wt, Stecker:1978ah, Kelner:2006tc} and photons\ \cite{Stecker:1978ah, Mucke:1999yb, Hummer:2010vx} to produce high-energy pions.  When they decay, they produce TeV--PeV neutrinos via $\pi^+ \to \mu^+ + \nu_\mu$, followed by $\mu^+ \to \bar{\nu}_\mu + \nu_e + e^+$, and their charge-conjugated processes.  Thus, neutrinos are nominally expected to be produced with flavor ratios $\left( N_e:N_\mu:N_\tau \right) = \left( 1:2:0 \right)_{\rm S}$, with $N_\alpha$ the sum of $\nu_\alpha$ and $\bar{\nu}_\alpha$.  Because at these energies it is difficult to disentangle the relative contribution of $\nu$ and $\bar{\nu}$ in neutrino telescopes, existing analyses typically assume that they contribute equally to the flux.  Thus, below, $\nu_\alpha$ refers to $\nu_\alpha + \bar{\nu}_\alpha$, unless otherwise indicated.  Interaction with matter in the sources likely does not modify the flavor ratios after production\ \cite{Mena:2006eq, Razzaque:2009kq, Sahu:2010ap, Varela:2014mma, Xiao:2015gea}.  

Other production mechanisms may affect the flavor composition; we highlight two possibilities.  First, if the muons from pion decay significantly lose energy before decaying, \eg, by synchrotron radiation in a strong magnetic field, neutrinos born from muon decay have lower energies.  In this ``muon-damped'' scenario, the high-energy flavor ratios are $(0:1:0)_{\rm S}$.  Second, neutrons co-produced with pions beta-decay into $\bar{\nu}_e$, yielding $(1:0:0)_{\rm S}$.  Yet, these neutrinos are $\sim$100 times less energetic than those made in pion decays.  Throughout, we use  the three physically motivated scenarios --- full pion decay, muon damping, neutron decay --- as benchmarks.

Production of $\nu_\tau$ is expected to be strongly suppressed, since it would require producing rare mesons, like $D_s^\pm$.  Below, we explore the full breadth of production mechanisms --- including those with large $\nu_\tau$ content --- and the scenario that enforces no $\nu_\tau$ production.

The flavor ratios might evolve with energy\ \cite{Kashti:2005qa, Lipari:2007su, Hummer:2010vx, Hummer:2011ms, Bustamante:2015waa}.  In the main text, we assume that they are measured in a single energy bin, so that any flavor evolution is hidden.  This is the current experimental status\ \cite{Aartsen:2015ivb, Aartsen:2015knd}.  However, future neutrino analyses will allow to study the flavor composition in multiple high-energy bins; see the Supp.\ Mat.\ for the case of IceCube-Gen2.


\smallskip
{\bf Neutrino oscillations.---}   Because a neutrino of a given flavor $\nu_\alpha$ is a superposition of neutrino mass eigenstates $\nu_i$ ($i = 1, 2, 3$), it can change flavor as it propagates.  The connection between the flavor and mass bases is represented by the Pontecorvo-Maki-Nakagawa-Sakata (PMNS) unitary mixing matrix $\matr{U}$.  Following convention, we parametrize it in terms of three mixing angles, $\theta_{12}$, $\theta_{23}$, and $\theta_{13}$, and one CP-violation phase, $\delta_{\rm CP}$, whose values are measured in numerous oscillation experiments.

For TeV--PeV astrophysical neutrinos, the probability $P_{\alpha\beta}$ of the flavor transition $\nu_\alpha \to \nu_\beta$ oscillates rapidly.  Because of the energy spread of neutrinos and the limited energy resolution of detectors\ \cite{Aartsen:2013vja}, flavor oscillations average out and the probability is\ \cite{Pakvasa:2008nx} $P_{\alpha\beta} = \sum_{i=1}^3 \lvert U_{\alpha i} \rvert^2 \lvert U_{\beta i} \rvert^2$, where $U_{\alpha i}$ are elements of the PMNS matrix.  Thus, the flavor ratios at Earth are $f_{\alpha,\oplus} = \sum_{\beta=e,\mu,\tau} P_{\beta\alpha} f_{\beta,{\rm S}}$.  If neutrinos are produced in the full pion decay chain and the probability is evaluated at the best-fit values of the mixing parameters, the flavor ratios at Earth are close to $\left( 1:1:1 \right)_\oplus$; this is the nominal expectation.  Flavor ratios can be used to probe fundamental neutrino physics, though we do not explore this possibility here; see, \eg, \Refs\ \cite{Esmaili:2009dz, Barenboim:2003jm, Bustamante:2010nq, Xu:2014via,  Arguelles:2015dca, Bustamante:2015waa, Shoemaker:2015qul, Bustamante:2016ciw, Brdar:2016thq, deSalas:2016svi, Rasmussen:2017ert, Bustamante:2018mzu, Ahlers:2018yom}.

\begin{figure*}[t!]
 \centering
 \includegraphics[width=\columnwidth]{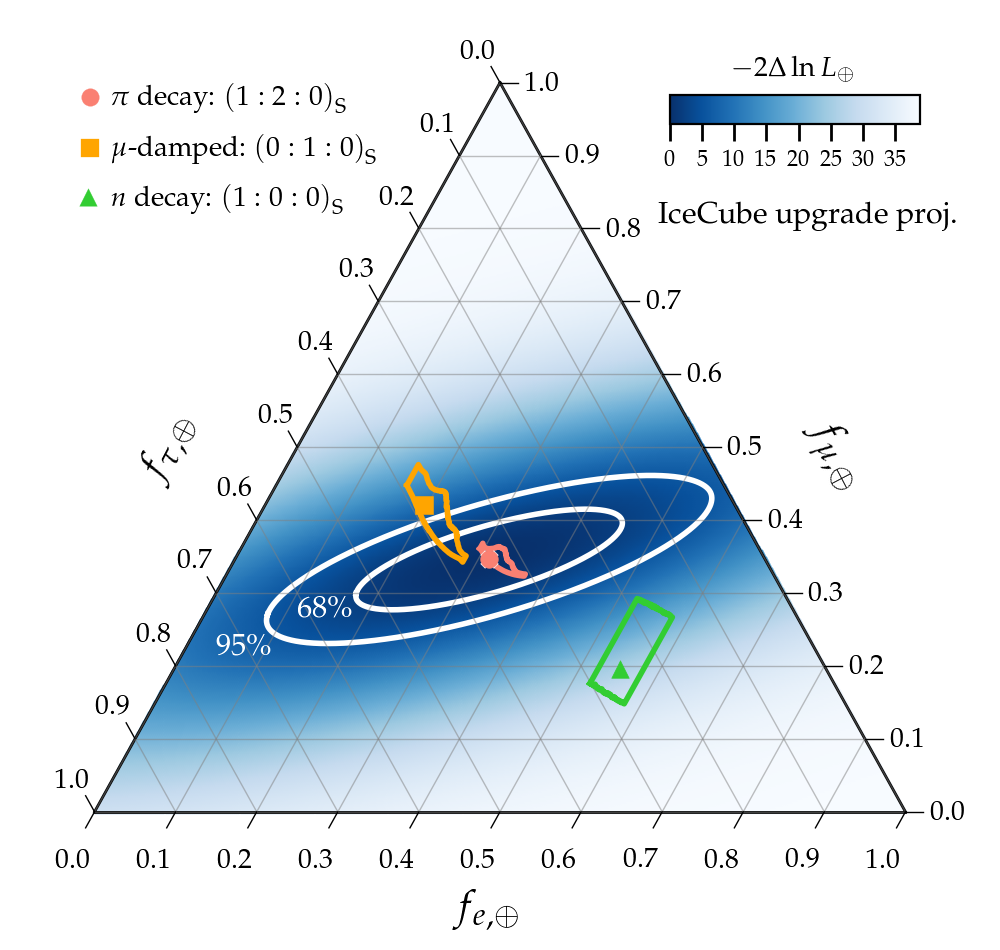}
 \includegraphics[width=\columnwidth]{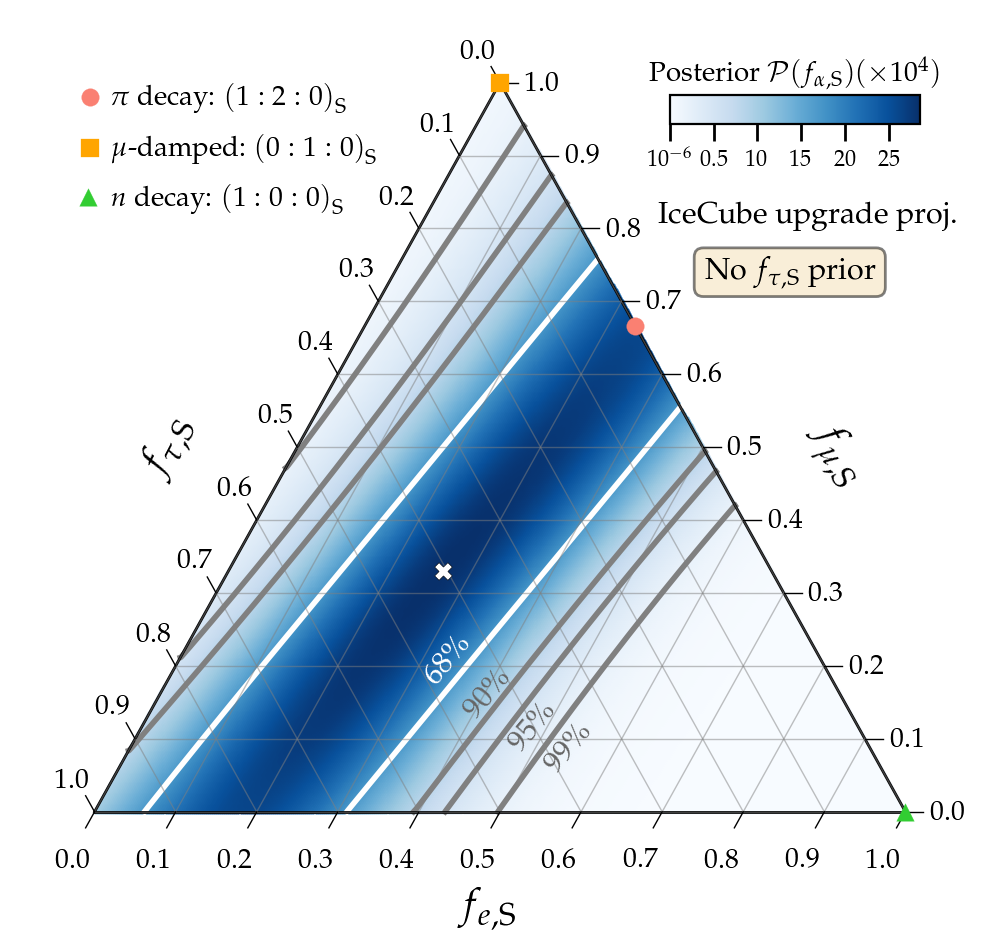}
 \caption{\label{fig:ic_upg}{\it Left:}  Same as \figu{ic_2015}, left, but showing the projected flavor sensitivity of the IceCube upgrade, approximated from \Ref\ \cite{blot_talk_tevpa_2018}.  {\it Right:}  Same as \figu{ic_2015}, right, but showing the projected performance of the IceCube upgrade in inferring $f_{\alpha, \oplus}$.}
\end{figure*}


\smallskip
{\bf Mixing parameters.---}  Presently, $\theta_{12}$ and $\theta_{13}$ are known at $1\sigma$ to within 2\%, $\theta_{23}$ to within 8\%, and $\delta_{\rm CP}$ to within 20\%.  This translates into uncertainties of around 20\% in transition probabilities, which we account for below.
For our analysis, we adopt the allowed ranges of mixing parameters from the NuFit 3.2 global fit to oscillation data\ \cite{Esteban:2016qun, NuFit_Jan2018}, assuming normal neutrino mass ordering ($s_{ij} \equiv \sin \theta_{ij}$): $s_{12}^2 = 0.307^{+0.013}_{-0.012}$, $s_{23}^2 = 0.538^{+0.033}_{-0.069}$, $s_{13}^2 = (2.206 \pm 0.075) \cdot 10^{-2}$, and $\delta_{\rm CP} = (234^{+43}_{-31})^\circ$.  The phase $\delta_{\rm CP}$ has only marginal influence on the flavor composition at Earth, since it appears in the flavor-transition probabilities suppressed by two or four powers of the small angle $s_{13}$.  Using inverted ordering does not affect our conclusions; we show this in the Supp.\ Mat.

We account for the uncertainties in the mixing parameters via their probability density functions (PDFs) $\mathcal{P}$.  
For each parameter in the set $\boldsymbol\theta \equiv \left( s_{12}, s_{23}, s_{13}, \delta_{\rm CP} \right)$, we approximate its PDF as a normal distribution with the mean and standard deviation computed, respectively, from the best-fit and largest $1\sigma$ error above.
This is justified because the $\Delta \chi^2$ curves that represent their uncertainties are nearly symmetric around the best-fit values\ \cite{Esteban:2016qun, NuFit_Jan2018}.  Future implementations of our proposed method could use refined PDFs built directly from the $\Delta \chi^2$ curves.

Figure \ref{fig:ic_2015}, left, shows for the three benchmark production scenarios that varying the mixing parameters within their $3\sigma$ uncertainties results in small allowed regions of flavor ratios at Earth.  Yet, these uncertainties, small though they seem, are an important limiting factor when reconstructing flavor ratios at the sources.


\smallskip
{\bf Measuring flavor at Earth.---}  IceCube is presently the largest, most sensitive detector of high-energy neutrinos\ \cite{Ahlers:2018fkn}.  It instruments a giga-ton of clear Antarctic ice with an array of strings of photomultipliers\ \cite{Aartsen:2016nxy}.  When a high-energy neutrino interacts with a nucleon in the vicinity of the detector, final-state charged particles initiate particle showers that radiate Cherenkov light, which is collected by the photomultipliers.
In the TeV--PeV neutrino energy range, IceCube detects two types of neutrino-induced event topologies: elongated tracks from high-energy muons --- initiated mainly by interactions of $\nu_\mu$ --- and approximately spherical showers from electromagnetic and hadronic cascades --- initiated by all flavors, but predominantly by $\nu_e$ and $\nu_\tau$. Comparing the relative numbers of tracks and showers yields the flavor ratios $f_{\alpha,\oplus}$\ \cite{Mena:2014sja,Aartsen:2015ivb, Aartsen:2015knd,Palladino:2015vna, Palomares-Ruiz:2015mka, Vincent:2016nut, Aartsen:2018vez}.  
At higher energies, flavor-specific event topologies due to $\bar{\nu}_e$\ \cite{Glashow:1960zz, Anchordoqui:2004eb, Bhattacharya:2011qu, Barger:2014iua, Palladino:2015vna} and $\nu_\tau$\ \cite{Learned:1994wg, Beacom:2003nh, Bugaev:2003sw}, already hinted at by current data\ \cite{Talk_Stachurska_TeVPA_2018, Talk_Lu_UHECR_2018}, might improve flavor and $\nu$ {\it vs.} $\bar{\nu}$ discrimination.

Figure \ref{fig:ic_2015}, left, shows the latest published IceCube flavor measurements\ \cite{Aartsen:2015knd}, covering energies between 25~TeV and 2.8~PeV, expressed via the likelihood function $\mathcal{L}_\oplus(f_{e, \oplus}, f_{\mu, \oplus})$.  Since precise IceCube likelihood data are not public, here and below we approximate present and future IceCube likelihood functions as two-dimensional normal distributions in $f_{e, \oplus}$ and $f_{\mu, \oplus}$; unitarity demands $f_{\tau, \oplus} = 1 - f_{e, \oplus} - f_{\mu, \oplus}$.  Because of the low statistics of present analyses, the confidence regions are broad.  Because $\nu_e$- and $\nu_\tau$-initiated showers are similar, they are currently not distinguished from one another on an event-by-event basis; see, however, \Ref\ \cite{Li:2016kra}.  This is why the iso-contours in \figu{ic_2015} are approximately horizontal, aligned with a direction of constant $f_{e, \oplus} + f_{\tau, \oplus}$. The degeneracy is weakly broken because $\nu_\tau$ interactions create muon tracks 17\% of the time, unlike $\nu_e$.  The best fit is at $(0.49:0.51:0)_\oplus$, about $1\sigma$ away from the nominal expectation.  Later, we consider projected improvements in flavor measurement.

IceCube measures the flavor composition of the diffuse flux of high-energy astrophysical neutrinos.  Since the diffuse flux is the aggregated contribution of multiple sources --- which possibly emit neutrinos with different flavor ratios --- the flavor ratios $f_{\alpha, \oplus}$ measured by IceCube are the average of all sources.  By extension, so are the flavor ratios at the sources $f_{\alpha, {\rm S}}$ that we infer below.


\smallskip
{\bf Inferring flavor at the sources.---}  For a given test choice of flavor ratios at the sources, we assess its compatibility with the data by computing an associated Bayesian posterior probability density that factors in the uncertainties in mixing parameters --- via their PDFs --- and the detector performance in measuring flavor ratios --- via the likelihood $\mathcal{L}_\oplus$.  The posterior probability density of $f_{\alpha, {\rm S}}$ being the flavor ratios at the sources is
\begin{equation*}\label{equ:likelihood_source}
 \mathcal{P} \left( f_{\alpha,{\rm S}} \right)
 \equiv
 \int {\rm d} \boldsymbol\theta
 \frac{\mathcal{P}(\boldsymbol\theta)}{ \mathcal{N}\!\left( \boldsymbol\theta \right) }\mathcal{L}_\oplus \left[ f_{e,\oplus}( f_{\alpha,{\rm S}}, \boldsymbol\theta ), f_{\mu,\oplus}( f_{\alpha,{\rm S}}, \boldsymbol\theta ) \right]\;,
\end{equation*}
where $\mathcal{P}(\boldsymbol\theta)~\equiv~\mathcal{P}(s_{12}) \mathcal{P}(s_{23}) \mathcal{P}(s_{13}) \mathcal{P}(\delta_{\rm CP})$ are the PDFs of the mixing parameters and
\begin{equation*}
 \mathcal{N}\!\left( \boldsymbol\theta \right) 
 \equiv
 \int\limits_0^1 {\rm d} f_{e, {\rm S}} \!\!\!\!\!\int\limits_0^{1-f_{e, {\rm S}}}\!\!\!\!\! {\rm d} f_{\mu, {\rm S}} 
 \mathcal{L}_\oplus \left[ f_{e,\oplus}( f_{\alpha,{\rm S}}, \boldsymbol\theta ), f_{\mu,\oplus}( f_{\alpha,{\rm S}}, \boldsymbol\theta ) \right]
\end{equation*}
is a normalization constant.

We compute the posterior of all possible values of $f_{\alpha, {\rm S}}$.  After that, we calculate credible intervals of $f_{\alpha, {\rm S}}$ by integrating the posterior, starting from its global maximum, down to the desired level, \eg, 68\%, 90\%, 95\%, or 99\%.

A previous analysis\ \cite{Mena:2014sja} also inferred the flavor composition at the sources, using early IceCube data.  However, unlike the present analysis, it did not account for uncertainties in the mixing parameters, which are crucial for the interpretation of the data.


\smallskip
{\bf Present results.---}  Figure \ref{fig:ic_2015}, right, shows the posterior of all possible flavor ratios at the sources, computed based on the latest published IceCube flavor measurements\ \cite{Aartsen:2015knd}, \figu{ic_2015}, left.  The maximum-posterior composition is $(0:0.7:0.3)_{\rm S}$, and compositions with low $f_{e, {\rm S}}$ and high $f_{\mu, {\rm S}}$ are preferred.  This is a consequence of the current preference for low $f_{\tau,\oplus}$ in the IceCube likelihood, which maps compositions at the sources close to the $f_{e, {\rm S}} = 0$ axis into compositions at Earth with a high likelihood value.  

Among the benchmark scenarios included in \figu{ic_2015}, production via pion decay with muon damping is allowed at the 68\%~credible level (Cr.L.), the full pion decay chain is slightly less favored, and neutron decay is in tension with the data, since it is allowed only at more than the 99\%~Cr.L.  Later, we explore how this changes if future IceCube flavor likelihood functions are centered instead on a nearly equi-flavor composition.

Because the production of $\nu_\tau$ should be suppressed, next we supplement our method by introducing the prior $f_{\tau, {\rm S}} = 0$.  With it, the posterior becomes a function of only $f_{e, {\rm S}}$, since $f_{\mu, {\rm S}} = 1-f_{e, {\rm S}}$. Figure \ref{fig:posterior_source_no_tau}, bottom, shows the resulting one-dimensional posterior: the maximum-posterior composition and 68\%~credible interval is $f_{e, {\rm S}} = 0.04^{+0.27}_{-0.04}$.


\smallskip
{\bf Future prospects.---}  Larger event samples, advances in flavor-tagging, and reduced uncertainties in mixing parameters will significantly improve how well flavor ratios at the sources are inferred.  Below, we estimate prospects for the IceCube upgrade\ \cite{blot_talk_tevpa_2018} --- to be built in the near future, with 7 new in-fill detector strings --- and for 15 years of running of the planned IceCube-Gen2\ \cite{Aartsen:2014njl} --- with 5--7 times the effective area.

Figure \ref{fig:ic_upg}, left, shows the projected flavor likelihood of the IceCube upgrade\ \cite{blot_talk_tevpa_2018}.  Unlike the present-day likelihood, the projected one is maximum, by design, at the nominal expectation of $\left( 0.31, 0.35, 0.34 \right)_\oplus$, \ie, the nearly equi-flavor composition at Earth expected from production by the full pion decay chain, $\left( 1:2:0 \right)_{\rm S}$, computed using the present best-fit values of the mixing parameters. The same is true for IceCube-Gen2, though with flavor contours 5 times tighter; see the Supp.\ Mat.  

Figure \ref{fig:ic_upg}, right, shows the posterior computed based on the projected likelihood of the IceCube upgrade, \figu{ic_upg}, left, without applying any prior on $f_{\tau, {\rm S}}$.  The maximum posterior is at $\left( 0.25:0.33:0.42 \right)_{\rm S}$ --- not far from flavor equipartition --- even though the IceCube likelihood was designed assuming $\left( 1:2:0 \right)_{\rm S}$.  The reason behind this is subtle, but consistent with our Bayesian approach; we detail it in the Supp.\ Mat.  By imposing again the prior $f_{\tau, {\rm S}} = 0$, we are able to sidetrack this subtlety and recover $\left( 1:2:0 \right)_{\rm S}$ as the maximum-posterior composition.

Figure \ref{fig:posterior_source_no_tau} shows projections for the posterior assuming $f_{\tau, {\rm S}} = 0$ in the IceCube upgrade and IceCube-Gen2.  For IceCube-Gen2, we assume that the mixing parameters will be known with negligible uncertainties compared to the width of the likelihood.
 
Assuming that neutrino production indeed occurs via pion decay, \figu{posterior_source_no_tau} shows that, in the near future, the IceCube upgrade could disfavor the muon-damped scenario at the 95\%~Cr.L. and the neutron-decay scenario at more than the 99\%~Cr.L.  The uncertainty on $f_{e,{\rm S}}$ is expected to shrink by a factor of 2.5.  Later, in IceCube-Gen2, the uncertainty could be up to 10 times smaller than today, allowing us to single out the composition from pion decay and rule out alternatives.  The Supp.\ Mat.\ shows that, if production includes muon damping, the performance of IceCube-Gen2 will be comparable to \figu{posterior_source_no_tau}.  These studies could measure or constrain the average magnetic field strength in neutrino sources\ \cite{Baerwald:2011ee}.  In reality, analyses performed at the time of operation of IceCube-Gen2 will need to factor in the finite, but small expected uncertainties in the mixing parameters.


\smallskip
{\bf Summary and outlook.---}  The study of the flavor composition of high-energy astrophysical neutrinos can help to identify their unknown production mechanism.  We have introduced a method to infer the flavor composition at the neutrino sources based on measurements of the composition at Earth and on the allowed ranges of values of the neutrino mixing parameters.  We hope that our results encourage neutrino-telescope collaborations, present and future, to perform further analyses in this direction.

Based on published IceCube data, we found that production of neutrinos via the decay of high-energy pions is compatible with data at the 90\% credible level (Cr.L.), while the scenario where intermediate muons in the pion decay chain cool in strong magnetic fields is slightly favored, at the 68\%~Cr.L..  Production via neutron decay is strongly disfavored, at more than 99\%~Cr.L. 

In the future, the IceCube upgrade and extension, IceCube-Gen2, should be capable of singling out the production mechanism and firmly excluding alternatives.  This will require synergy between astrophysical-neutrino experiments --- to reduce uncertainties in flavor measurements --- and oscillation experiments --- to reduce uncertainties in neutrino mixing parameters.  On both fronts, progress is ongoing, with promising prospects.


\smallskip
{\bf Acknowledgements.}  We thank Carlos Arg\"uelles, Siqiao Mu, and Sergio Palomares-Ruiz for useful discussion.  MA and MB are supported by the Danmarks Grundforskningsfond Grant 1041811001 and \textsc{Villum Fonden} (projects no.~18994 and no.~13164, respectively).  This work used resources provided by the High Performance Computing Center at the University of Copenhagen.



%


\newpage
\clearpage

\appendix


\onecolumngrid

\begin{center}
 \large
 Supplemental Material for\\
 \smallskip
 {\it Inferring the flavor of high-energy astrophysical neutrinos at their sources}
\end{center}

\bigskip

\twocolumngrid


\section{The maximum in projected posteriors}

In \figu{ic_upg} in the main text we show that the posterior computed based on the projected likelihood of the IceCube upgrade, without applying any prior on $f_{\tau, {\rm S}}$, is not maximum at $\left( 1:2:0 \right)_{\rm S}$, even though the underlying likelihood was built to be maximum at $f_\oplus^\ast \equiv \left( 0.31, 0.35, 0.34 \right)_\oplus$, the composition at Earth expected from production by the full pion decay, computed using the present best-fit values of the mixing parameters.  Below, we explain the reason behind this.

First, consider the ideal case where the values of the mixing parameters are known with perfect accuracy and are equal to their present best-fit values.  In this case, $f_\oplus^\ast$ maps back onto $\left( 1:2:0 \right)_{\rm S}$.  As a result, because $f_\oplus^\ast$ is the point of maximum IceCube likelihood, $\left( 1:2:0 \right)_{\rm S}$ receives the maximum posterior.  This is what we see in our projections for IceCube-Gen2, in \figu{posterior_source_ic_gen2}, left.

Now consider the realistic case that includes uncertainties in the mixing parameters.
In this case, the best-fit values of the mixing parameters that are required to regress from $f_\oplus^\ast$ to $\left( 1:2:0 \right)_{\rm S}$ enter the calculation of the posterior weighed down by their PDFs.  This dilutes the high likelihood associated to $f_\oplus^\ast$ and reduces the posterior of $\left( 1:2:0 \right)_{\rm S}$.  In contrast, $\left( 1:1:1 \right)_\oplus$ maps back onto $\left( 1:1:1 \right)_{\rm S}$ for any values of the mixing parameters.  Because of this, the likelihood of $\left( 1:1:1 \right)_\oplus$ enters the calculation of the posterior not weighed down by the PDFs; nearby points are weighed down only mildly.  As a result, points near $\left( 1:1:1 \right)_{\rm S}$ have the highest posterior.  This is what we see in our projection for the IceCube upgrade, \figu{ic_upg}, right.

In \figu{posterior_source_no_tau} in the main text, we show that imposing the prior $f_{\tau, {\rm S}} = 0$ sidetracks this subtlety. \\

\vfill
\newpage


\section{IceCube-Gen2 flavor projections}

Here we present further projections of the performance of IceCube-Gen2 in inferring flavor ratios at the sources.  

Figure \ref{fig:likelihood_exp_ic_gen2} shows the projected flavor likelihood of IceCube-Gen2, assuming production via the full pion decay chain --- below neutrino energies of 1 PeV --- and muon damping --- above 1 PeV.  The projections that we use here approximate those shown in \Ref\ \cite{kowalski_NOW_2018} as two-dimensional normal distributions, like we did for IceCube in the main text.

Figure \ref{fig:posterior_source_ic_gen2} shows the corresponding posterior probability density of flavor ratios at the sources, projected for IceCube-Gen2.  We show the posterior without any prior in the scenario of production via the full pion decay chain, and the posterior using the $f_{\tau, {\rm S}} = 0$ prior for both production scenarios.

With enough events detected across a wide range of energies, our method could be applied to IceCube-Gen2 data to look for evidence of the transition from production via the full pion decay chain to production with additional muon damping\ \cite{kowalski_NOW_2018}.


\section{Inverted mass ordering}

In the main text, we obtained results by using probability density functions $\mathcal{P}$ of the mixing parameters built under the assumption of normal neutrino mass ordering.  In this appendix we show selected results using probability density functions of the mixing parameters built under the assumption of inverted mass ordering.  In this case, the best-fit values and $1\sigma$ uncertainties of the mixing parameters are, extracted from the NuFit 3.2 global fit\ \cite{Esteban:2016qun, NuFit_Jan2018}, are ($s_{ij} \equiv \sin \theta_{ij}$): $s_{12}^2 = 0.307^{+0.013}_{-0.012}$, $s_{23}^2 = 0.554^{+0.023}_{-0.033}$, $s_{13}^2 = (2.227 \pm 0.074) \cdot 10^{-2}$, and $\delta_{\rm CP} = (278^{+26}_{-29})^\circ$. 

Figure \ref{fig:posterior_source_no_tau_ih} shows the posterior probability density of $f_{e, {\rm S}}$ under the prior $f_{\tau, {\rm S}} = 0$, for inverted mass ordering.  The plot is similar to \figu{posterior_source_no_tau}.  This illustrates that our conclusions in the main text are largely independent of the choice of mass ordering.

\vfill

\setcounter{figure}{0}
\renewcommand{\thefigure}{B\arabic{figure}}
\begin{figure*}[!t]
 \centering
 \includegraphics[width=0.497\textwidth]{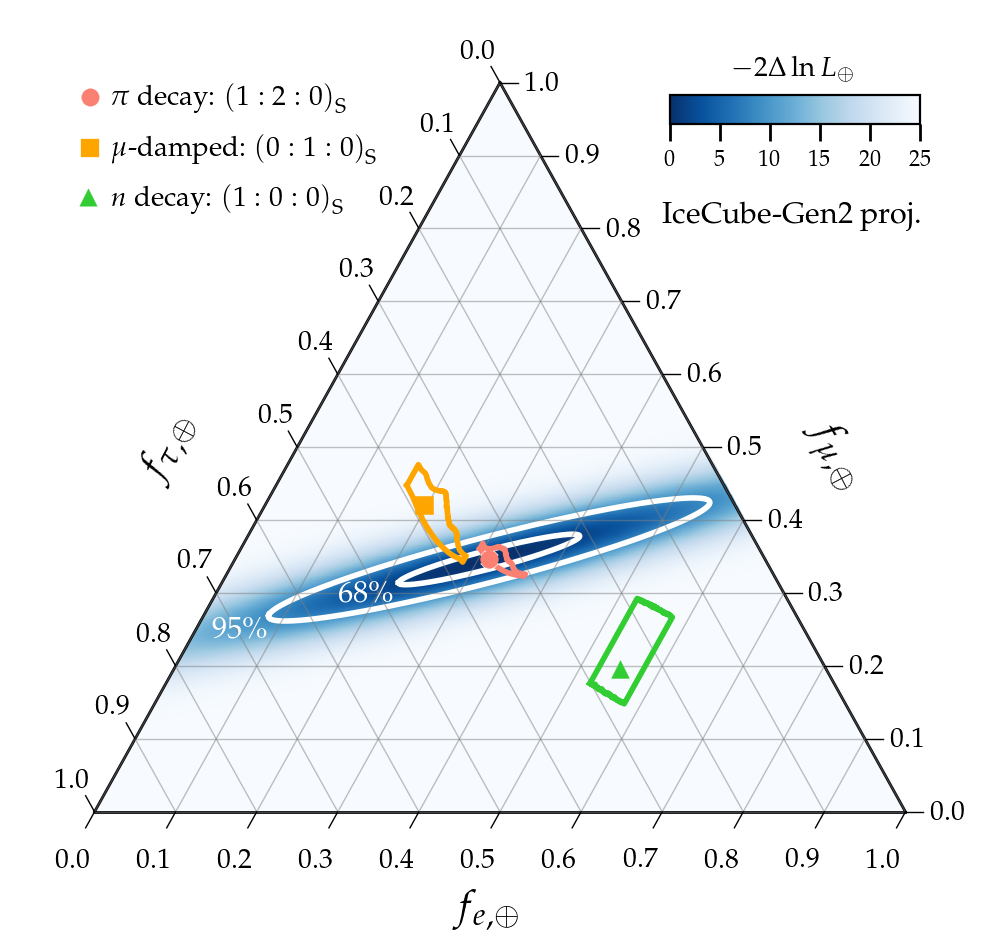}
 \includegraphics[width=0.497\textwidth]{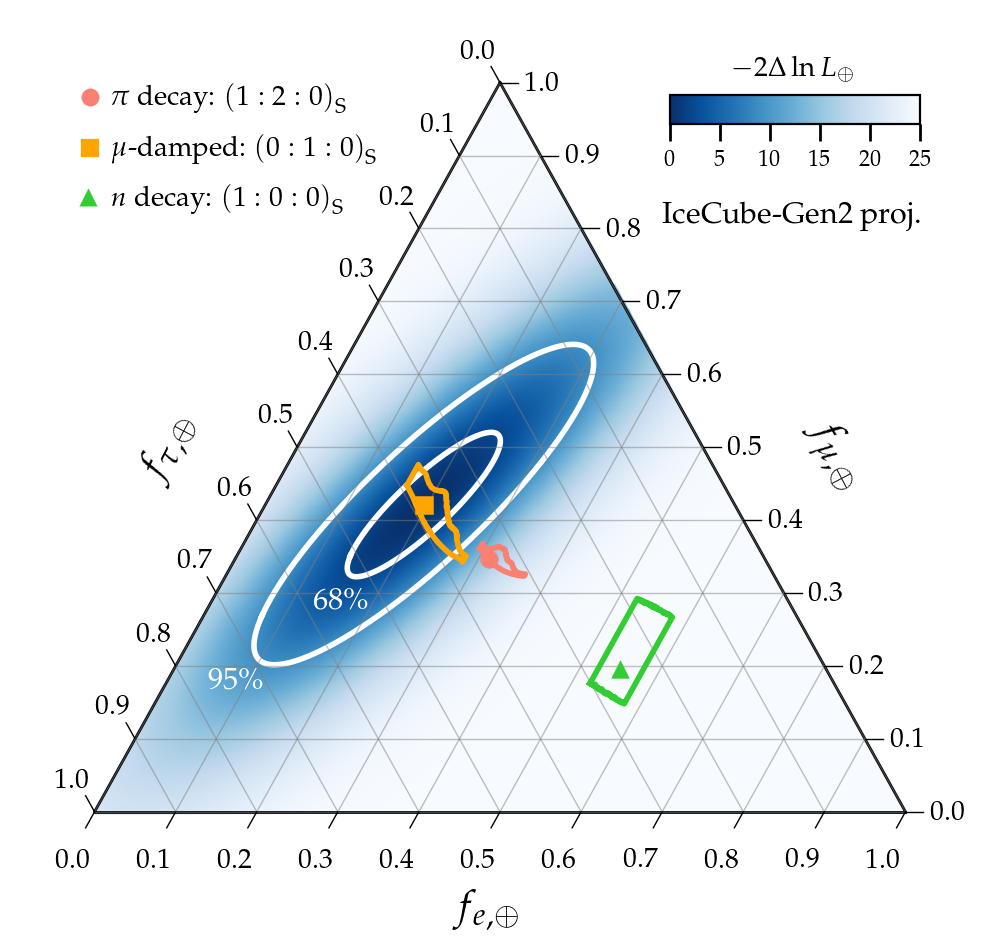}
 \caption{\label{fig:likelihood_exp_ic_gen2}Same as \figu{ic_2015}, left, but showing instead the estimated flavor performance of IceCube-Gen2.  {\it Left:} Centered on the composition due to full pion decay, $\left( 1:2:0 \right)_{\rm S}$\ \cite{kowalski_NOW_2018}.  {\it Right:} Centered on the composition due to pion decay with muon damping, $\left( 0:1:0 \right)_{\rm S}$\ \cite{kowalski_NOW_2018}.}
\end{figure*}

\begin{figure*}[b!]
 \centering
 \includegraphics[width=0.497\textwidth]{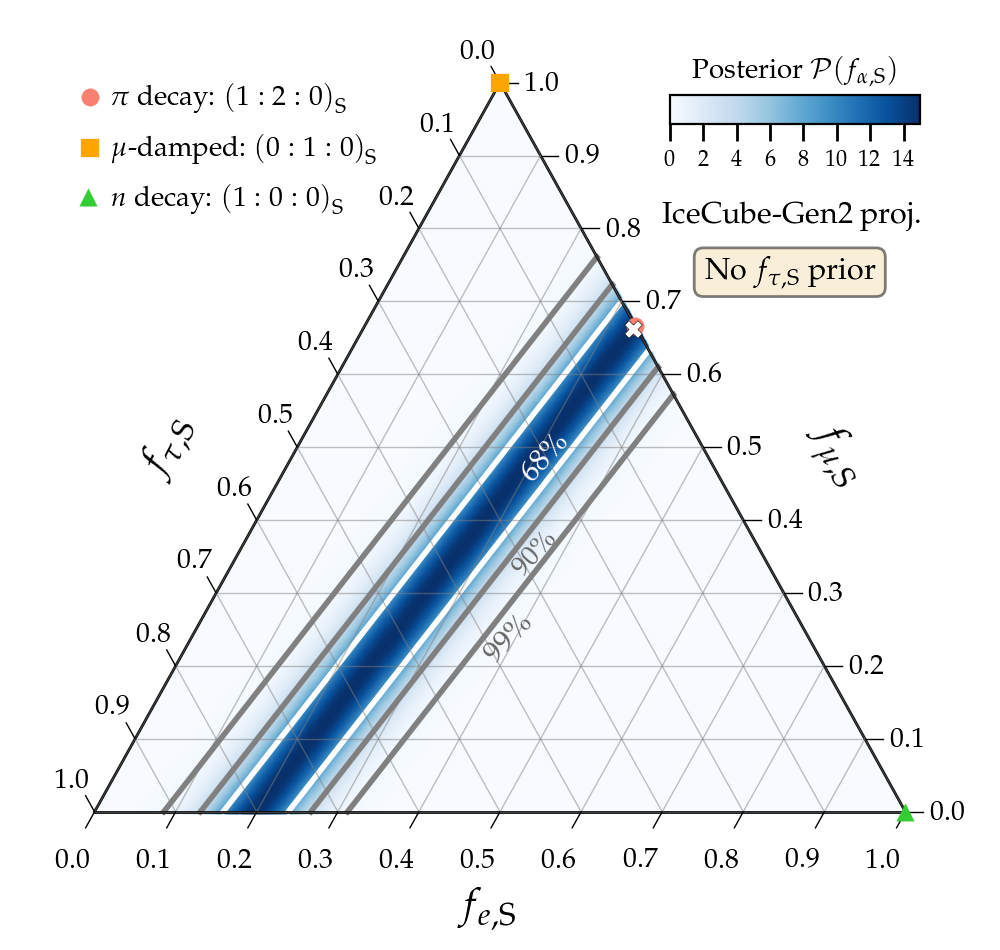} 
 \includegraphics[width=0.497\textwidth]{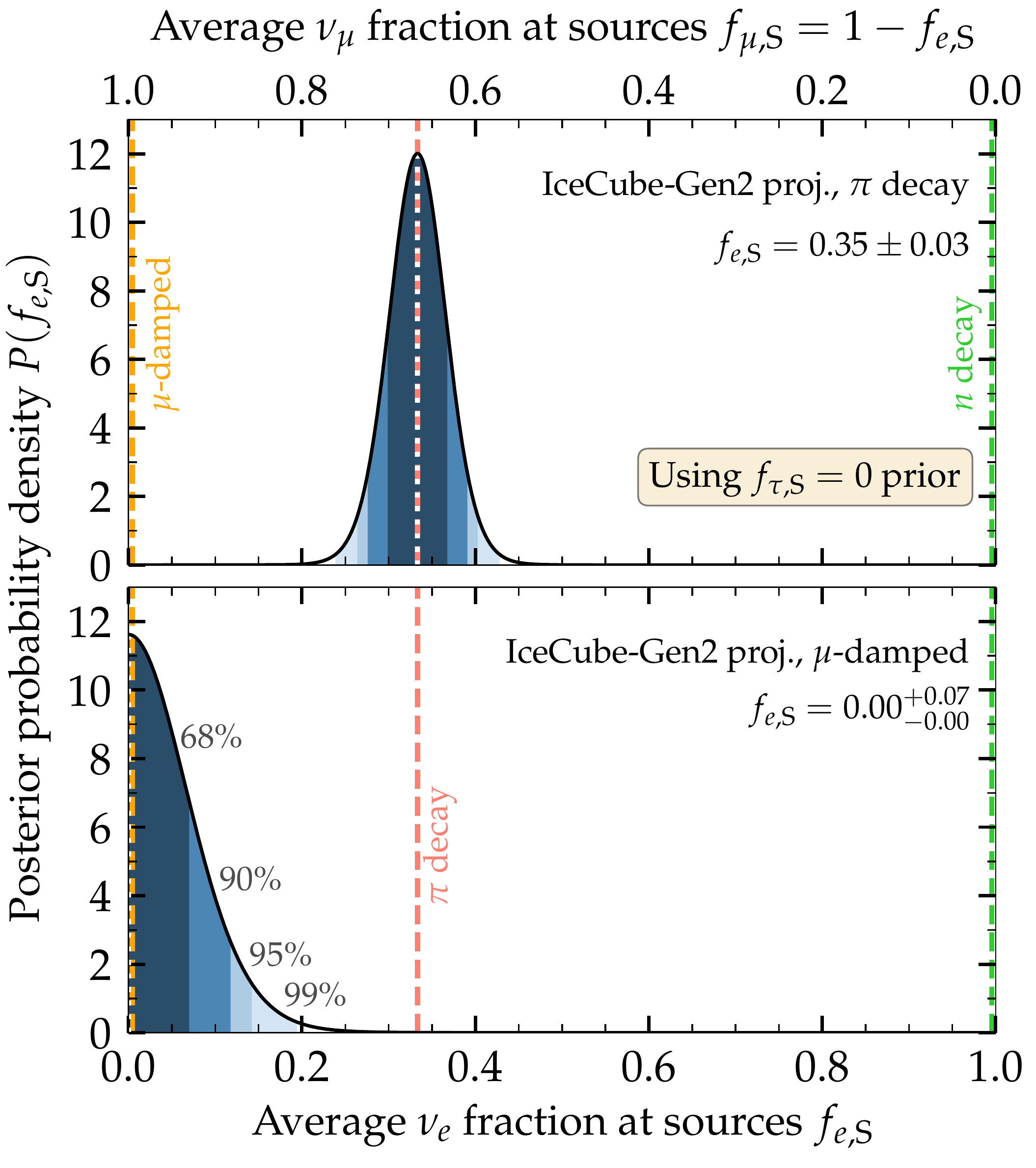}
 \caption{\label{fig:posterior_source_ic_gen2}{\it Left:} Same as \figu{ic_2015}, right, but showing instead the estimated flavor performance of IceCube-Gen2, assuming neutrino production by the full pion decay chain, \ie, using the likelihood from the left panel of \figu{likelihood_exp_ic_gen2} and no prior on $f_{\tau, {\rm S}}$.  {\it Right:} Enforcing a prior of $f_{\tau, {\rm S}} = 0$ and using the projected IceCube-Gen2 performance estimated for neutrino production via the full pion decay chain ({\it top}) and pion decay with muon damping ({\it bottom}), \ie, using the likelihood from the left and right panels of \figu{likelihood_exp_ic_gen2}, respectively.}
\end{figure*}

\setcounter{figure}{0}
\renewcommand{\thefigure}{C\arabic{figure}}
\begin{figure*}[t!]
 \centering
 \includegraphics[width=0.497\textwidth]{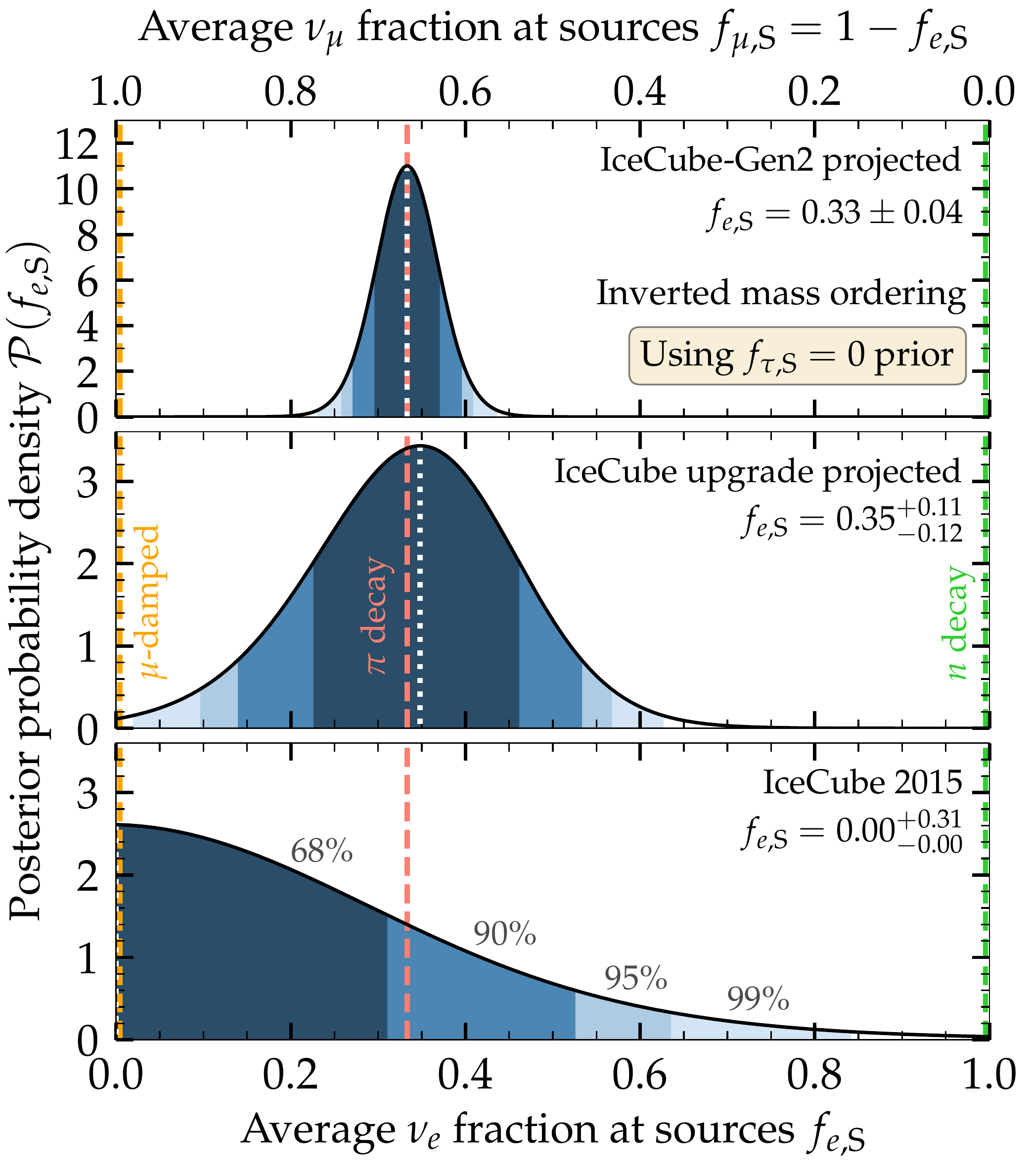}
 \caption{\label{fig:posterior_source_no_tau_ih}Same as \figu{posterior_source_no_tau}, but generated using the uncertainty ranges of mixing parameters derived under the assumption of an inverted neutrino mass ordering \cite{Esteban:2016qun, NuFit_Jan2018}.}
\end{figure*}

\end{document}